\documentclass[fleqn,usenatbib]{mnras}

\usepackage{newtxtext,newtxmath}
\usepackage[T1]{fontenc}

\DeclareRobustCommand{\VAN}[3]{#2}
\let\VANthebibliography\thebibliography
\def\thebibliography{\DeclareRobustCommand{\VAN}[3]{##3}\VANthebibliography}

\usepackage{graphicx}	% Including figure files
\usepackage{amsmath}	% Advanced maths commands
\usepackage{siunitx}
\DeclareSIUnit\year{yr}
\DeclareSIUnit\parsec{pc}
\DeclareSIUnit\ev{eV}
\DeclareSIUnit\solarmass{\textup{M}_\odot}

\title{Temperature asymmetry in the Milky Way’s hot circumgalactic medium induced by the Magellanic Clouds}
\author[A. Oprea et al.]{
Alexandru~Oprea,$^{1}$
Filippo~Fraternali,$^{1}$\thanks{fraternali@astro.rug.nl}
Else~Starkenburg,$^{1}$
Thor~Tepper-Garcia,$^{2}$
Joss~Bland-Hawthorn,$^{2}$
\\
$^{1}$Kapteyn Astronomical Institute, University of Groningen, Postbus 800, 9700 AV Groningen, The Netherlands\\
$^{2}$Sydney Institute for Astronomy, School of Physics, A28, The University of Sydney, NSW 2006, Australia
}
    
\date{Accepted 2026 February 10. Received 2026 January 27; in original form 2025 November 28}

\pubyear{2025}

\begin{document}
\label{firstpage}
\pagerange{\pageref{firstpage}--\pageref{lastpage}}
\maketitle

\begin{abstract}
The Milky Way is surrounded by a hot diffuse circumgalactic medium (CGM) with temperatures of millions of degrees. 
Recent X-ray observations with the eROSITA satellite discovered a significant temperature asymmetry of this hot CGM, with the southern hemisphere being on average hotter than the northern one by a relative difference of $\Delta T/T \approx 12\%$, where $T$ is averaged over the entire CGM. 
In this Letter, we investigate whether the passage of the Magellanic Clouds can be responsible for this asymmetry by means of a hydrodynamical/$N$-body simulation.
In the simulation, the Magellanic Clouds induce a relative motion of the Milky Way's disc of up to $40\,\unit{\kilo\meter\per\second}$.
This motion leads to compression of the CGM gas in the southern hemisphere, resulting in an overall temperature increase in that region.
We estimate a south-north temperature difference of $\Delta T/T \approx 13\%$–$20\%$, consistent with the observations.
We find that this temperature asymmetry is a recent phenomenon that began $\sim 100\, {\rm Myr}$ ago.
\end{abstract}

\begin{keywords}
Galaxy: halo --
(galaxies:) Magellanic Clouds -- 
galaxies: interactions --
hydrodynamics
\end{keywords}

%%%%%%%%%%%%%%%%%%%%%%%%%%%%%%%%%%%%%%%%%%%%%%%%%%

%%%%%%%%%%%%%%%%% BODY OF PAPER %%%%%%%%%%%%%%%%%%

\section{Introduction}

The Milky Way (MW) is surrounded by a diffuse medium of hot gas, often referred to as the hot circumgalactic medium (CGM) or Galactic corona. 
The existence of such a hot CGM was first postulated by \cite{Spitzer1956} as a required medium for the confinement of high-velocity clouds \citep[HVCs][]{1997ARA&A..35..217W} and was further supported by the observation that all satellites within the MW virial radius of $\approx230 \, \unit{\kilo\parsec}$ are devoid of gas \citep{CFN2020, 2021ApJ...913...53P}, except for the more massive Large and Small Magellanic Clouds (LMC and SMC, respectively). 
More directly, the hot CGM can be revealed by X-ray emission and absorption features. 
In particular, \cite{Miller&Bregman2015} analysed O\,{\small VII} and O\,{\small VIII} emission-line observations in $\approx 650$ sightlines and found a roughly spherical medium at a temperature of $\approx 2\times 10^6\,{\rm K}$ with a total mass of $4-5 \times 10^{10}\,{M_{\odot}}$, if extrapolated to the virial radius.
The hot CGM also appears to rotate \citep{Hodges-Kluck+2016}.

Recently, the eROSITA All-sky Survey has enabled mapping of the MW's hot CGM in the medium-energy X-ray regime, up to $10 \, \unit{\kilo\electronvolt}$, using O\, {\small VII} and O\,{\small VIII} emission lines with better angular and spectral resolution than previously possible \citep{2021A&A...647A...1P}. 
Using these data, \cite{erosita} have built a detailed (pseudo-)temperature map of the MW's CGM. 
They found that the temperature distribution, when looking at regions on scales of $2^{\circ}-20^{\circ}$, appears homogeneous and in the range $2.2- 2.5 \times 10^6 \,\unit{\kelvin}$. However, when the north and south galactic hemispheres are compared, significant temperature variations are present. They reported $\Delta T/T \approx 12\%$, on scales of tens of degrees, with higher temperatures in the southern hemisphere. 
The cause of this asymmetry remains unclear.
\citet{erosita} attribute it to an intrinsic difference within the large-scale hot CGM, of unknown origin. 
However, it is relevant to note that there can also be a degeneracy with the local hot bubble \citep{2024A&A...690A.399Y}.
 
It is well known that the distribution of satellites around the MW is asymmetric \citep{2012AJ....144....4M}. 
Most notably, the two largest and most massive systems, the aforementioned LMC and SMC, are found in the southern Galactic hemisphere. 
This interacting pair of galaxies is currently in orbit around the MW at 50$-$60$\,\unit{\kilo\parsec}$ distance \citep{Pietrzynski+2013,Graczyk+2014}. 
As first suggested by \cite{first_infall}, they are likely to be on their first passage around the Galaxy, although a second passage cannot be ruled out \citep[e.g.][]{2024MNRAS.527..437V}. 
The Galactocentric velocity of the satellites has been estimated to be in the range $300 - 400 \, \unit{\kilo\meter\per\second}$ \citep[][]{Kallivayalil2013, 2016ApJ...832L..23V, 2018A&A...616A..12G}. 

The total (virial) mass of the LMC, estimated at $\approx 1.8 \times 10^{11} \,\unit{\solarmass}$ from globular clusters kinematics \citep[][]{2024ApJ...963...84W}, is significant compared to the MW's virial mass \citep[$1.3 \pm 0.3 \times 10^{12} \, \unit{\solarmass,}$][]{2016ARA&A..54..529B} and its infall is expected to have a non-negligible influence on the MW's surroundings in several ways. 
Multiple stellar stream orbits can only be explained when the LMC's gravitational potential is taken into account\citep[e.g.,][]{sag_stream2, Erkal2018,orph_stream}. The LMC's perturbing force also provides a possible explanation for the warp of the MW disc \citep[e.g.,][]{2018MNRAS.473.1218L,Binney2024}. 

%\par\vspace{0pt}%
\begin{figure*}
    \includegraphics[width=\textwidth]{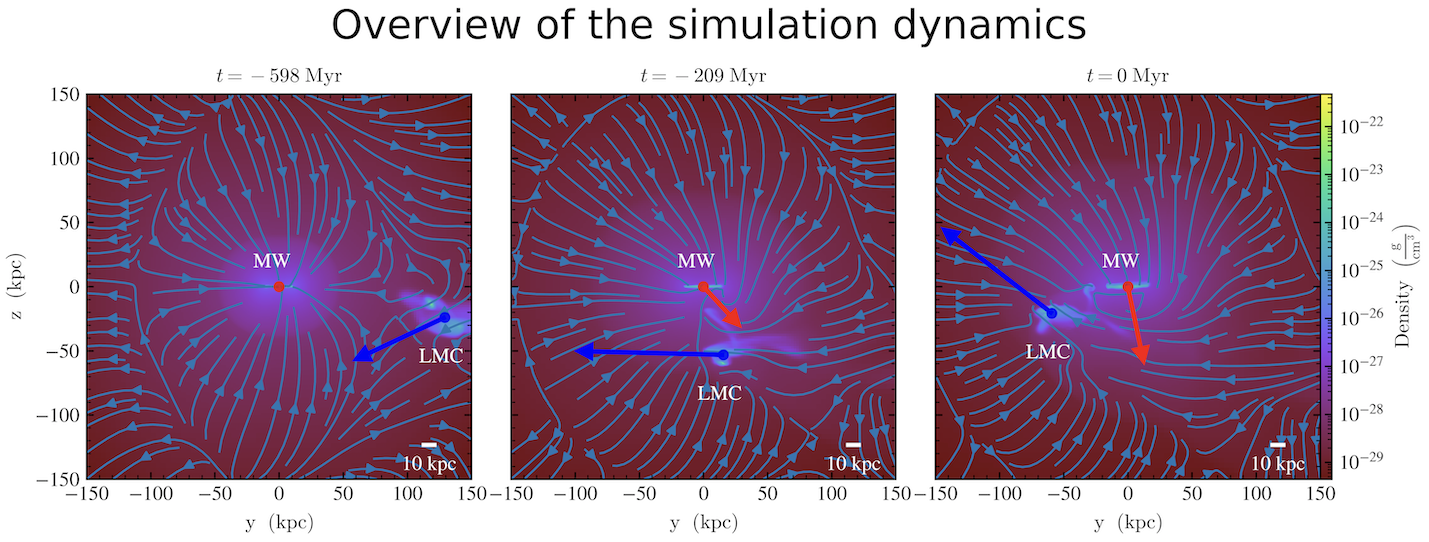}
    \caption{Simulations snapshots of the projected gas at epochs: $t=600 \,\unit{\mega\year}$ ago (left panel), $t=200 \,\unit{\mega\year}$ ago (centre) and at the present time (right panel). 
    Streamlines are overlaid to show the projected motion of the hot CGM gas. 
    The two main objects are marked, the LMC system (including the SMC) and the MW disc (MW), which is kept in the centre in this projection. 
    Their velocity vectors are represented by blue and red arrows, respectively. For clarity, the red arrows scale is $1\,\unit{\kilo\parsec}$ per $\unit{\kilo\meter\per\second}$ while for the blue arrows $1\,\unit{\kilo\parsec}$ corresponds to $3 \,\unit{\kilo\meter\per\second}$. 
    In the left panel, the MW disc is missing its arrow, as its velocity is negligible at that time. 
    }
    \label{fig:cartoon}
\end{figure*}

Motivated by the improved, and significantly increased, total mass estimate for the LMC at the time \citep{Kallivayalil2013}, \cite{anditmoves} showed that the approximation of an inertial Galactocentric reference frame is invalid, requiring a common centre of mass when modelling orbits. These authors estimated that the MW's centre of mass would have been displaced by as much as $30 \, \unit{\kilo\parsec}$ with velocities of up to $75 \, \unit{\kilo\meter\per\second}$ \citep[see also][]{Petersen2020}. As the passage of the LMC is expected to affect the stellar and dark matter halo of the Galaxy both kinematically and spatially \citep[a quantified prediction can be found in e.g.,][]{2019ApJ...884...51G}, many observational studies have attempted to measure and validate these effects \citep[e.g.,][]{Belokurov2019,2021NatAs...5..251P,Conroy2021,Amarante2024,Cavieres2025,Bystrom2025}.

In this Letter, we investigate whether the Magellanic Clouds could also be responsible for the observed temperature asymmetry in the MW's hot CGM. We address this question through the analysis of an $N$-body/hydrodynamical simulation originally designed to study the interaction between the MW and the Magellanic Clouds and the formation of the Magellanic stream with a focus on the survival of the Leading Arm \citep{Thor_2019}. 
In the following sections, we describe this simulation and show why it is suitable for the investigation of this problem. We then quantify the effect of the LMC on the MW's CGM as reflected in this simulation and compare it to the eROSITA observations.

\section{Method}

\subsection{Description of the simulation}

The simulation used in this paper was originally designed to investigate the formation of the Magellanic Stream and the Leading Arm of the LMC-SMC system \citep{Thor_2019}. 
It was performed using RAMSES \citep{RAMSES}, an $N$-body/hydrodynamical code using adaptive mesh refinement. 
The original study presents six models of the LMC-SMC system, of which only four include a Galactic CGM, the focus of our investigation. 
All four models include a live dark matter halo for both the MW and the LMC-SMC system. 
The difference between the models lies in the rotation speed of the CGM and the inclusion of a magnetic field for the non-rotating CGM. 
The two rotating CGM models, denoted 'Slow corona' and 'Fast corona', have rotation speeds in the plane of the Galaxy of $\approx 70 \,{\rm km\,s^{-1}}$, and $\approx 135 \,\unit{km\,s^{-1}}$, respectively.
Here, we use the 'Slow corona' model, which is in better agreement with cosmologically motivated models of the hot CGM \citep{c44f98dd56ca407bacf43bc8b9649688} at the radii where the interaction with the LMC-SMC system takes place. 

In the selected simulation, the MW's halo is modelled as a live dark-matter component with an NFW profile \citep{1997ApJ...490..493N} having a virial mass of $\simeq 1.5 \times 10^{12} \,\unit{\solarmass}$ and a scale radius of $20 \,\unit{\kilo\parsec}$. 
The CGM follows the profile as the dark matter with the same structural parameters and a mass of $\simeq 3 \times 10^{10} \,\unit{\solarmass}$ within $250 \,\unit{\kilo\parsec}$.
We note that this profile is more centrally concentrated than $\beta$-models, typically used in observations \citep[e.g.][]{Miller&Bregman2015}.
The rotation speed of the CGM follows a profile that decreases both radially and vertically.
Both the LMC and the SMC are modelled as dark matter halos hosting a stellar and a gas component. 
These halos follow NFW profiles with scale lengths of $13 \,\unit{\kilo\parsec}$ and $3.8 \,\unit{\kilo\parsec}$, respectively.
The virial masses are $1.75 \times 10^{11} \,\unit{\solarmass}$ and $0.15 \times 10^{11} \,\unit{\solarmass}$ for LMC and SMC, respectively. 
The gas and stellar components are rotating discs with declining exponential density profiles, with integrated masses significantly lower than for the dark matter \citep{Thor_2019}.

Following \cite{2018ApJ...857..101P}, the Magellanic Clouds were evolved as a binary pair in isolation for $\sim 6 \,\unit{\giga\year}$ with initial conditions such that when they are placed in the MW potential, the gas distribution and the distance between the LMC and SMC resemble observational constraints. The placement of the binary system in the MW potential is done by mapping the clouds' coordinates in the galactic reference frame in such a way that the LMC-SMC system matches today's observation after the simulation is run for $\approx 1 \,\unit{\giga\year}$.

\begin{figure*}
    \includegraphics[width=\textwidth]{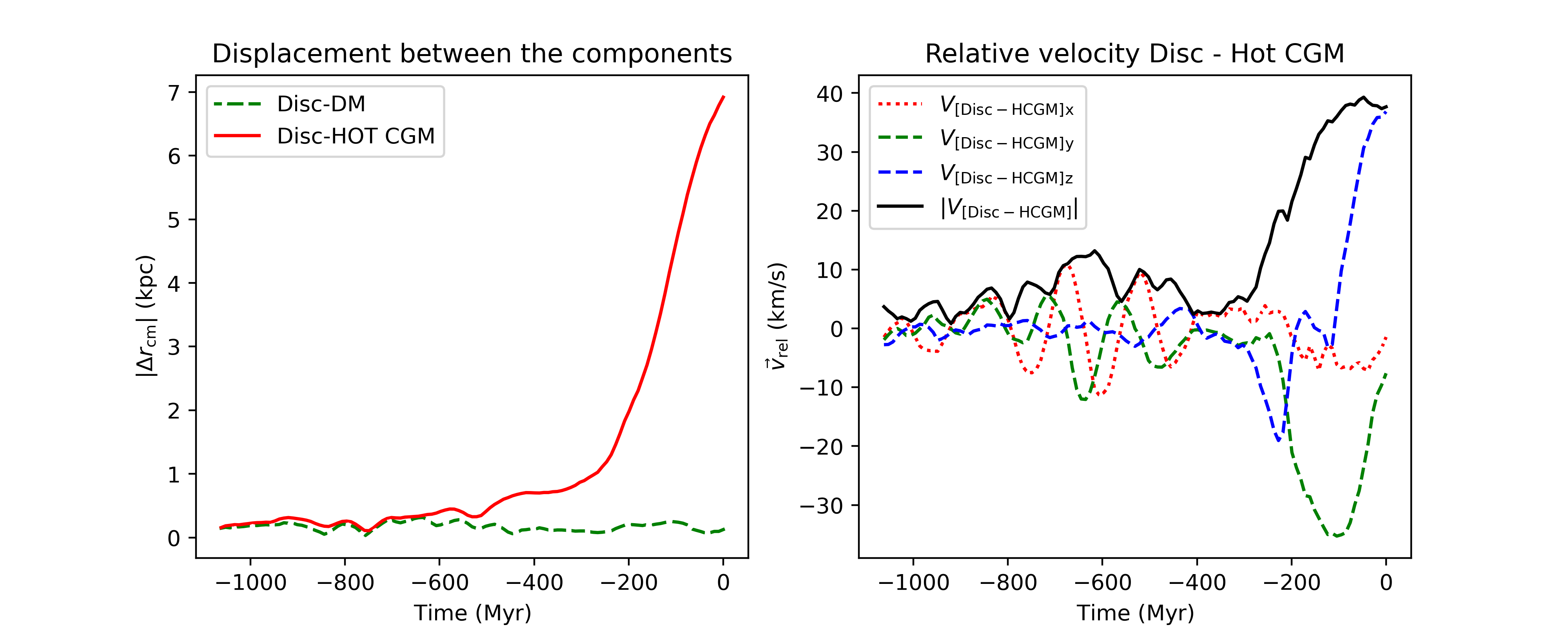}
    \caption{Displacement and kinematics of hot CGM, disc, and inner dark matter halo components. The $x$-axis represents the simulation time referenced to the present day. {\em Left panel}: the relative distance between the centres of mass of the disc and the hot CGM (red) and between the centres of mass of the disc and the inner dark matter halo (green). 
    {\em Right panel}: relative velocities between disc and hot CGM. The colors red, green, and blue show the $x$, $y$, and $z$ components of the velocity vector with respect to the simulation axis, while the black curve presents its total magnitude.}
    \label{fig:kinematics}
\end{figure*}

The simulation initially did not contain a disc of gas or stars in the centre of the MW potential.
However, due to the cooling of the CGM, a cold ($T<10^5 \,\unit{\kelvin}$) disc forms quickly. 
This can be seen as a thin blue horizontal feature in the centre of each panel in Fig.\ \ref{fig:cartoon} labelled "MW". 
Due to the fact that star formation is not included in these simulations, this disc does not have a final stellar component. 
However, as we argue below, for the purpose of our study, the simulated disc is a good proxy for the actual MW disc and the conclusions drawn from this simulation would be unchanged if a stellar disc were present. 

First, despite not containing a stellar component, the baryonic disc's structural and rotational properties are similar to those of the MW. 
In particular, at the present day, the cold-gas disc has a total mass of $\approx 10^{10} \,\unit{\solarmass}$, a radius of $17.5 \,\unit{\kilo\parsec}$ and a roughly constant rotation speed of $\simeq 220 \,\unit{\kilo\meter\per\s}$.
These properties are comparable to those of the MW disc \citep{2016IAUS..317..266G, 2016ARA&A..54..529B}. 
Second and most crucially, as we show in Section \ref{sec:kinematics}, our cold-gas disc moves together with the inner (within $r\approx 40\,{\rm kpc}$) dark matter halo, feeling the gravitational pull of the LMC-SMC system, as is expected for the actual disc of the MW \citep[see e.g., Fig. 7 of][]{2019ApJ...884...51G}.

\subsection{Analysis of the different components}

Figure 1 shows the gas density snapshot corresponding to the present day, i.e., the snapshot in which the LMC position and velocity most closely match observations. 
For our analysis, we define two phases of the gas, hot and cold, using a temperature threshold of $10^5 \,\unit{\kelvin}$, which is typically used since the gas at this temperature is thermally unstable \citep[e.g.,][]{cooling}. 
As a consequence, we have three main components whose properties we follow as a function of time: the MW's cold disc, the hot CGM and the dark matter halo.
To consistently measure displacements and relative velocities among these three components, we define specific methods and apply them to all snapshots. 

For the gas disc, we select all the cold ($T<10^{5}\,{\rm K}$) gas within a box around the simulation centre of size $50 \times 45 \times 28 \, \unit{\kilo\parsec}^3$. 
The box size was chosen to encompass the displacement of the disc over the course of the simulation, while being smaller than the distance to the LMC to ensure exclusion of its gas. 
The centre of the disc is then taken as the centre of mass of the cold gas in this region. 
The position of the disc's centre of mass will further be used in the remainder of this work as a reference frame for different computations and is shown in Fig.\ \ref{fig:cartoon} as a red circle.

For the hot gas, we define the centre of the CGM as the centre of mass of the hot gas over a spherical volume of radius $40 \,\unit{\kilo\parsec}$ placed at the simulation centre. We verify that any choice of value for this radius between $20 \,\unit{\kilo\parsec}$ and $100 \,\unit{\kilo\parsec}$ does not significantly alter our results. 
The choice of $40 \,\unit{\kilo\parsec}$ allows the inclusion of hot gas up to the LMC while not including gas too far from the disc, which is insignificant to the results. 
The dark matter halo receives the same treatment as the hot gas, but uses only dark matter particles tagged for the MW.

We choose to compute the velocities of the gaseous components as the ratios between the displacement in position and the time interval instead of using the average particle velocities. 
This was done to average over random motions present in the hot gas that could overpower the bulk motion of the components.

\section{Results}

\subsection{Kinematics of the three components}
\label{sec:kinematics}

After the Magellanic Clouds are introduced in the simulation, the dark matter, hot CGM, and the disc start to move together under their influence.
The left panel of Fig.\ \ref{fig:kinematics} shows that in the first $500 \,\unit{\mega\year}$, the relative displacement between the components is negligible.
At those times, the disc and the hot CGM display a relative oscillatory motion in the disc plane with amplitudes not exceeding $10 \,\unit{\km\per\second}$, as shown by the right panel of Fig.\ \ref{fig:kinematics}.

%\label{section_graphs}
\begin{figure*}
    \includegraphics[width=\textwidth, trim={0 6cm 0 6cm},clip]{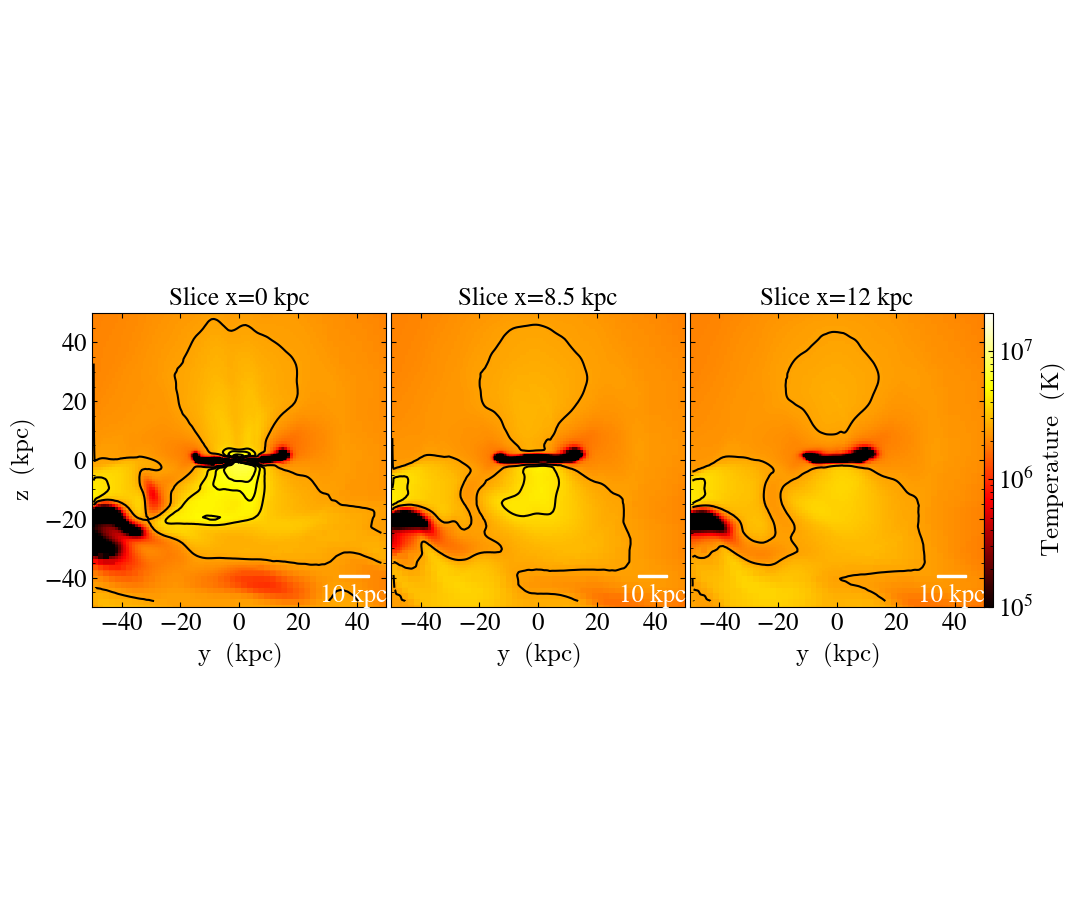}
    \caption{Cross-section slice plots in the $yz$ plane of the gas temperature above $10^5 \,\unit{\kelvin}$ extracted from the present-time snapshot of the simulation. The black regions correspond to temperatures $\leq 10^5 \,\unit{\kelvin}$ and show the MW disc (centre) and the LMC disc (bottom left). 
    The left plot shows a slice centred at the MW's nucleus, the middle plot a slice at $8.5 \,\unit{\kilo\parsec}$ from the centre, and the right plot a slice at $12 \,\unit{\kilo\parsec}$ from the centre along the positive $x$ axis; all including only one layer of cells in the $xy$ plane. 
    The isothermal contours are plotted for $2.5, 4, 5, 6\times10^6\,\unit{\kelvin}$.}
    \label{fig:temperature}
\end{figure*}

The behaviour of the three components starts to change when the LMC is at a distance of $d\approx150 \,\unit{\kilo\parsec}$ (see Fig.\ \ref{fig:cartoon}), which corresponds to a simulation time $t\approx-500 \,\unit{\mega\year}$ (with 0 being the present time). 
At this point, the cold-gas disc and the hot CGM begin to separate, while the disc's kinematics remains tightly coupled to that of the inner MW's dark matter halo (Fig.\ \ref{fig:kinematics}, left).
As the LMC moves further away from the disc plane, the decoupling of the CGM from the disc becomes more significant. 
Around $t=-200 \,\unit{\mega\year}$, the disc begins to move more rapidly with respect to the CGM, initially in the $y$ axis towards the direction of the LMC and eventually vertically acquiring a large $z$-velocity component (Fig.\ \ref{fig:kinematics}, right). 
At this time, the LMC is passing below the disc, inducing an acceleration in the vertical direction. 
Note that Fig.\ \ref{fig:cartoon} shows motions in the simulation frame, where both the disc and the hot gas move towards negative $z$. 
However, in the middle panel at $t=-209 \,\unit{\mega\year}$ the hot gas moves faster in this direction than the disc, whereas at present (right panel) this behaviour is reversed. 
This results in the sign change in the relative velocity of the $z$ component visible in Fig.\ \ref{fig:kinematics} (right).

At the present time, the LMC is placed on the opposite side (along the $y$-axis) with respect to where it started, resulting in the cancellation of the disc's y-movement. Thus, the relative disc-CGM motion at present is predominantly vertical with a magnitude of $\simeq 40 \,\unit{\km\per\second}$, which is consistent with the values found in the literature, as we discuss in Section \ref{sec:discussion_motion}. 

We can interpret the separation between cold and hot gas as the result of an interplay between gravitational and hydrodynamical effects.  
The cold-gas disc and the inner dark matter halo move together as one component under the gravitational action of the LMC. 
This, in turn, makes the kinematics of the disc in this simulation both realistic and reflective of the actual kinematic behaviour of the MW disc.
The hot CGM, instead, is composed of gas in hydrostatic equilibrium and feels its own pressure; thus, its global movement is delayed. 
The result is that the hot gas experiences a "sloshing", similar to the hot gas in galaxy clusters \citep{2018PASJ...70....9H, 2024MNRAS.529..563R}. 
This difference leads to a separation between the disc and the hot gas, which is crucial for what follows.

\subsection{Temperature asymmetry}
In order to study the impact of the relative motion of the MW disc with respect to the CGM, we fix the frame in the centre of the disc and view the CGM as we would see it from an observational perspective. To compare with the eROSITA observations and their striking north-south (N-S) asymmetry, we pay particular attention to any differences arising in the simulation's northern and southern hemispheres, at positive and negative $z$-coordinates, respectively. 

In the simulation, the initial temperature distribution when the LMC approaches the disc is symmetric with respect to the $xy$ plane, where the MW disc lies. 
At $t=-500 \, \unit{\mega\year}$, the average temperature in the north and south regions contained within $|z|<20\,\unit{\kilo\parsec}$ is $2.5 \times 10^6 \,\unit{\kelvin}$. 
Figure \ref{fig:temperature} shows the present-day temperature distribution in three different slices in the $yz$ plane at various distances from the MW centre. 
Clearly, a strong asymmetry in the gas temperature is present between the north and south.
The motion of the MW disc, in response to the passage of LMC, has resulted in a compression of the southern hot gas, increasing its temperature. 
Figure \ref{fig:temperature} shows that the effect is most significant along the central axis of the disc. 
In these regions, out to $z\approx -5\,\unit{\kilo\parsec}$, the temperature reaches $6 \times 10^6 \,\unit{\kelvin}$. At $x=8.5 \,\unit{\kilo\parsec}$ away from the centre, roughly the Solar Neighbourhood, the temperature in the same vertical region is $4 \times 10^6 \,\unit{\kelvin}$, also significantly higher than the average initial temperature of the CGM. 
We note that a similar compression effect is present in front of the LMC. 
However, the gas in that region does not appear to influence the southern CGM in proximity to the disc. 
We conclude that the heating effect is dominated by the MW disc's motion with respect to the hot CGM.

Figure \ref{fig:temperature} also shows that the temperature in the northern CGM does not change significantly, maintaining a value of $2.5 \times 10^6 \,\unit{\kelvin}$.
In that region, the present-day temperature closely follows the initial distribution, but with expanded isotherms.
Its topology has not changed in the last $200 \, \unit{\mega\year}$\footnote{A video with the evolution of the temperature distribution over the simulation's timescale can be accessed at [see additional material]}.
In this respect, we note that our simulation does not include feedback from star formation or the central black hole so the temperature of the CGM is fully determined by the initital conditions and the compression induced by the LMC.
  
From the above, it is clear that the decoupling between the MW disc and hot CGM that we found in the simulation could explain the asymmetry in the temperature of the north and south galactic hemispheres recently observed with eROSITA by \cite{erosita}. 
To quantify this, we selected two regions in the simulations at locations similar to those in the observations and find the relative temperature difference $\Delta T/T$. 
We fix the Sun at $8.5 \,\unit{\kilo\parsec}$ away from the centre and place it in the disc such that the LMC coordinates match the projected galactic coordinates. 
The regions we considered are two circles of radius $20^{\circ}$ centred at $(l, b) = (220, +50) $ and $(l, b) = (230, -50)$, to mimic the regions used in \cite{erosita}. 
In the simulation frame, we thus examine conic volumes in the gas.
We note that these regions are safely unaffected by the presence of the eROSITA bubbles in the central parts of the CGM \citep[see Fig.\ 9 of][]{erosita}. 
The magnitude of the resulting N-S asymmetry depends on the truncation distance of the conic volumes. 
Given that the distance to the line transitions in observations is not well constrained, we vary the truncation distance in the simulation and examine its effect. 
We take density-weighted averages of the gas temperature. 
Taking a truncation distance of $10 \,\unit{\kilo\parsec}$ or $40 \,\unit{\kilo\parsec}$ results in an asymmetry $\Delta T/T \approx 20 \%$.
A truncation at $30 \,\unit{\kilo\parsec}$ gives an asymmetry $\Delta T/T \approx 16 \%$, while at $20 \,\unit{\kilo\parsec}$ $\Delta T/T \approx 13 \%$. 
Overall, we find a significant asymmetry between the north and south hemispheres of the CGM at a level consistent with to the observed value.

\section{Discussion}
\label{section_discussion}

\subsection{Comparison with the literature}
\label{sec:discussion_motion}

Our results suggest that the passage of the LMC induced a vertical motion of the disc of the Galaxy with a magnitude of $40 \,\unit{\km\per\second}$. \cite{anditmoves} found in their analytic model that, under the action of the LMC, the centre of mass of the MW can be displaced as much as $30 \,\unit{\kilo\parsec}$ and acquire a velocity of up to $75 \,\unit{\km\per\second}$. 
Their model predicts that this displacement takes a short period of time that ranges from $300 \,\unit{\mega\year}$ to $500 \,\unit{\mega\year}$. 
Given that displacement and velocity depend on the mass chosen for the LMC and other components, our results match well within the uncertainties. 
The disc in our simulation starts to move significantly $\simeq 300 \,\unit{\mega\year}$ ago (Fig.\ \ref{fig:kinematics}, right), also in agreement.

A similar analysis has been performed on the stellar halo by \cite{2020MNRAS.494L..11P}. 
Using a $N$-body simulation, they predicted the motion of the stellar disc of $40 \,\unit{\km\per\second}$ with respect to the system barycentre under the influence of the LMC. 
Our results are consistent with their findings \citep[see also][]{Erkal+2021}. 
When we translate our results into Galactic coordinates, the disc moves in the direction $(l,b) = (79, -78)$ with respect to the CGM, and this direction broadly agrees with the general observed direction of the stellar disc movement relative to the distant outer stellar halo \citep{2021NatAs...5..251P}. 

We note that recently, \citet{Carr+2025} presented an $N$-body/hydrody-namical simulation of the interaction between LMC and the MW to study its impact on the MW's hot CGM.
This simulation differs substantially from ours in that it includes a CGM for LMC but no cold-gas disc for either LMC and the MW and no radiative cooling of the gas.
The authors briefly describe N-S asymmetries in their simulation, only noting a large-scale temperature asymmetry extending beyond the MW's virial radius and diametrically opposed to that observed by eROSITA. 
Such an asymmetry is unlikely to be observable because i) X-ray fluxes are dominated by the gas closer by and ii) it is primarily visible in the direction of LMC, while \citet{erosita} selected regions in the opposite direction.
By construction, \citet{Carr+2025}'s simulations do not have a MW cold disc and are, therefore, unable to capture the compression of the southern CGM that we report in this paper.
Their simulations, however, appear to produce some temperature enhancement in the Southern hemisphere (see their Fig.\ 4), likely arising from the compressive effect of the LMC's CGM.

\subsection{Implications for extraplanar cold gas clouds}

The region of the disc-halo interface (few to tens of ${\rm kpc}$ above and below the disc) of the MW and other disc galaxies is typically populated by gas clouds visible in HI emission \citep{Sancisi+2008}.
In the MW, these are called high- and intermediate-velocity clouds (HVCs and IVCs) \citep[e.g.][]{Wakker2004}. 
Together, they form an extraplanar gas layer, characterised by peculiar kinematics with respect to the disc \citep{Marasco&Fraternali2011}.
Warm ionised HVCs and IVCs are also observed, typically via absorption lines against quasars and halo stars, and share similar kinematics and localisation as their HI-emitting (denser) counterparts \citep[e.g.][]{Lehner+2022}.

HVCs and IVCs have long been known to show striking asymmetry, with the northern Galactic sky much more populated than the southern sky \citep{LAB,Putman+2012}.
This asymmetry has not been explained to date, but we may have some insights in light of our present findings. 
There is a consensus that IVCs originate via galactic fountains. 
However, HVCs may also be part of the same picture, if the ejected gas triggers cooling/condensation of part of the hot CGM \citep{Fraternali+2015, Fox+2023}.
Here, we found that the MW's disc has been moving toward (roughly) the Galactic south pole and with respect to the hot CGM in the last $\sim 100\,{\rm Myr}$, comparable to or less than the orbital times of fountain clouds \citep{2017ASSL..430..323F}.
Due to a combination of the displacement of the disc during their orbits and the drag of the hot CGM being lower in the Galactic north, it is natural to expect that fountain clouds would move to higher distances in the northern hemisphere, thus boosting the number of clouds there classified as IVCs and HVCs.
We plan to investigate this further in a future work.

\section{Conclusions}
We have used the $N$-body/hydrodynamical simulation of \cite{Thor_2019}, initially designed to study the Magellanic stream with a focus on the survival of the Leading Arm, to investigate the hydrodynamic effect of the Magellanic Clouds on the hot CGM surrounding the MW disc. 
Our main goal was to explore whether the passage of the Magellanic Clouds can play a role in the recently discovered N-S asymmetry in the temperature of the hot CGM.
Here we summarise our results.
\begin{itemize}
\item{
In the simulation, we identified three main components associated with the MW: the cold gas disc, the hot CGM and the dark matter halo.
The cold gas disc forms out of the cooling of the hot CGM and has properties similar to the actual MW disc. 
We followed the motion of these three components and found that the cold disc and the inner dark matter halo move together attracted by the Magellanic Clouds, while the hot CGM lags behind.
As a consequence, the cold disc acquires a relative velocity with respect to the hot CGM. This velocity is mostly vertical (toward the south pole) in the last $100 \, \unit{\mega\year}$ and reaches $\simeq 40\,\unit{\kilo\meter\per\second}$ at the present time, in agreement with other determinations in the literature. 
}
\item{
As a consequence of the disc motion, the southern hot CGM is compressed resulting in significant heating.  
In contrast, the northern gas is not disturbed and maintains the initial temperature distribution.   
Quantitatively, at the present time, the temperature close to the disc plane in the southern area is $4-6 \times 10^6 \,\unit{\kelvin}$ within the Solar circle, whereas the temperature in the north remains the initial $2.5 \times 10^6 \,\unit{\kelvin}$ on average. 
}
\item{
We mimic the observations of eROSITA from \citet{erosita}, by selecting two circular regions of radius $20^{\circ}$ centred at $(l, b) = (220, +50)$ and $(l, b) = (230, -50)$ to estimate the N-S asymmetry. 
Using a density-weighted average of the conic section of gas corresponding to the same regions in the simulation, we obtain an asymmetry in temperature $\Delta T/T$ between $13\%$ and $20\%$ depending on the distance probed, which well matches the value of $12\%$ in the observations. 
}
\end{itemize}

The above results show that the passage of the Magellanic Clouds can fully explain the N-S asymmetry observed by eROSITA in the temperature of the CGM.
We speculate that the motion of the MW disc with respect to the hot CGM, induced by the Magellanic Clouds, can also be responsible for the long-known N-S asymmetry in HCVs and IVCs.

\section{Acknowledgements}
We are grateful to an anonymous referee for a constructive report.
This work has received funding from the European Research Council (ERC) under the Horizon Europe research and innovation programme (Acronym: FLOWS, Grant number: 101096087 and Acronym: EARLYMW, Grant number: 101170507).
ES acknowledges funding through VIDI grant "Pushing Galactic Archaeology to its limits" (with project number VI.Vidi.193.093) which is funded by the Dutch Research Council (NWO).
This research has been partially funded from a Spinoza award by NWO (SPI 78-411).
The handling of the simulation data, both for computation and for plotting purposes, was done using the YT Python library YT \citep{YT}.

\section{Data availabity}
The simulation data used in this paper will be shared on reasonable request to the corresponding author.



\bibliographystyle{mnras}
\bibliography{ref} % if your bibtex file is called example.bib

%\begin{thebibliography}{99}
%\end{thebibliography}

%%%%%%%%%%%%%%%%%%%%%%%%%%%%%%%%%%%%%%%%%%%%%%%%%%

%%%%%%%%%%%%%%%%% APPENDICES %%%%%%%%%%%%%%%%%%%%%

%\appendix


%If you want to present additional material which would interrupt the flow of the main paper, it can be placed in an Appendix which appears after the list of references.

%%%%%%%%%%%%%%%%%%%%%%%%%%%%%%%%%%%%%%%%%%%%%%%%%%

% Don't change these lines
\bsp	% typesetting comment
\label{lastpage}
\end{document}